%% file: K3Pi.tex
\documentclass[preprint,12pt,superscriptaddress]{revtex4}  

\usepackage{graphics}
\usepackage{ulem}
\usepackage[usenames]{color}
\usepackage{epsfig}
\usepackage[usenames,dvipsnames]{xcolor}

\def\mevm{~MeV/$c^2$\/}

\def\gevm{~GeV/$c^2$\/}
\def\gevp{~GeV/$c$\/}

\def\rws{R^{}_{\textrm{ws}}}
\def\rawrws{R'_{\textrm{ws}}}
\def\dcs{D^{0}\rightarrow K^{+}\pi^{-}\pi^{+}\pi^{-}}
\def\cf{D^{0}\rightarrow K^{-}\pi^{+}\pi^{+}\pi^{-}}
\def\mk3p{M^{}_{K3\pi}}

\def\kbar{\overline{K}{}^{\,0}}
\def\dbar{\overline{D}{}^{\,0}}
\def\bbar{\overline{B}{}^{\,0}}
\def\bsbar{\overline{B}{}^{\,0}_s}

\def\kkbar{$K^0$-$\kbar$}
\def\ddbar{$D^0$-$\dbar$~}
\def\bbbar{$B^0$-$\bbar$~}
\def\bsbsbar{$B^0_s$-$\bsbar$~}

\begin{document}



\title{\large
\begin{flushright}
{\normalsize KEK Preprint 2013-27} \\
\vspace*{-0.17in}
{\normalsize UCHEP-13-02} 
\end{flushright}
\boldmath{
Measurement of the Wrong-Sign Decay 
$D^0 \rightarrow K^{+}\pi^{-}\pi^{+}\pi^{-}$
}}

\input{authors_pub398.tex}

\begin{abstract}

A measurement of the rate for the ``wrong-sign'' decay 
$D^{0}\rightarrow K^{+}\pi^{-}\pi^{+}\pi^{-}$ relative to 
that for the ``right-sign'' decay 
$D^{0}\rightarrow K^{-}\pi^{+}\pi^{+}\pi^{-}$ is presented.
Using 791 fb$^{-1}$ of data collected with the Belle detector,
we obtain a branching fraction ratio of
$\rws = [0.324 \pm 0.008 (\textrm{stat.}) \pm 0.007 (\textrm{sys.})]\%$.
Multiplying this ratio by the world average value for the
branching fraction ${\cal B}(D^0\rightarrow K^-\pi^+\pi^+\pi^-)$
gives a branching fraction
${\cal B}(D^{0}\rightarrow K^{+}\pi^{-}\pi^{+}\pi^{-}) = 
(2.61 \pm 0.06\,^{+0.09}_{-0.08}) \times 10^{-4}$.
\end{abstract}


\maketitle


Studies of mixing in neutral meson systems have had an important 
impact on the development of the Standard Model. Historically, 
mixing was first observed in the \kkbar~system~\cite{kkbar_mixing},
then later in the \bbbar system~\cite{bbbar_mixing}, and most 
recently in the 
\bsbsbar~\cite{bsbsbar_mixing} 
and \ddbar~\cite{belle_kk,babar_kp,LHCb_kp}~systems. Mixing 
in the \ddbar~system is strongly suppressed due to  
Cabibbo-Kobayashi-Maskawa (CKM)~\cite{ckm_matrix} 
matrix elements and the GIM mechanism~\cite{gim}. It has been measured 
using several methods~\cite{HFAG_preprint}, 
one of which compares the time-dependence of
``wrong-sign'' $D^{0}\rightarrow K^{+}\pi^{-}(X)$ decays to
that of ``right-sign'' $D^{0}\rightarrow K^{-}\pi^{+}(X)$ 
decays~\cite{babar_kp,belle_kp,babar_kpp,cdf_kp,LHCb_kp}.
Wrong-sign decays can occur either via a doubly 
Cabibbo-suppressed (DCS) amplitude such as 
$D^{0}\rightarrow K^{+}\pi^{-}(X)$ or via 
$D^{0}$ mixing to $\dbar$, followed by a Cabibbo-favored (CF) 
decay such as $\dbar \rightarrow K^{+}\pi^{-}(X)$.

In this report we present a measurement for the rate of 
the wrong-sign (WS) decay 
$\dcs$ relative to that of the right-sign (RS) decay $\cf$
using a data sample of 791~fb$^{-1}$~\cite{cc}.
Assuming negligible $CP$ violation, the ratio of decay
rates can be expressed as~\cite{Bergmann}
\begin{eqnarray} \label{eq:rws}
\rws & \equiv & \frac{\Gamma(\dcs)}{\Gamma(\cf)} \nonumber \\
 & =  & R_{D} + \alpha y^{\prime}\sqrt{R_{D}} + \frac{1}{2}(x^{\prime 2} + y^{\prime 2})\,,
\end{eqnarray}
where $R_{D}$ is the squared magnitude of the ratio of the 
DCS to CF amplitudes, $\alpha$ is a suppression factor that 
accounts for strong-phase variation over 
the phase space ($0 \le \alpha \le 1$)~\cite{babar_kpp},
and $x^{\prime}$ and $y^{\prime}$ are the mixing parameters 
$x \equiv \Delta m/\overline{\Gamma}$ and 
$y \equiv \Delta \Gamma/2\overline{\Gamma}$ rotated by the effective 
strong phase difference $\delta$ between  DCS and CF 
amplitudes: $x^{\prime} = x\cos\delta + y\sin\delta$ and
$y^{\prime} = y\cos\delta - x\sin\delta$.
The parameters $x$ and $y$ depend only on the mass difference 
($\Delta M$) and decay width difference ($\Delta\Gamma$) between 
the \ddbar mass eigenstates, and the mean decay width ($\overline\Gamma$).
The Belle collaboration has previously measured 
$\rws = [0.320 \pm 0.018 (\textrm{stat.})^{+0.018}_{-0.013}(\textrm{sys.})]\%$~\cite{belle_tian}.
The measurement presented here supersedes this previous result. 
We use an improved reconstruction code that has a higher reconstruction 
efficiency for low momentum tracks. The data used in this analysis 
corresponds to an integrated luminosity of 791 fb$^{-1}$ 
collected at or near the $\Upsilon(4S)$ resonance.


The data sample was collected by the Belle detector~\cite{belle_detector}
located at the KEKB asymmetric-energy $e^+e^-$ collider~\cite{kekb}.
The Belle detector is a 
large-solid-angle magnetic spectrometer consisting of a silicon vertex 
detector (SVD), a 50-layer central drift chamber (CDC), an array of 
aerogel threshold Cherenkov counters (ACC), a barrel-like 
arrangement of time-of-flight scintillation counters (TOF), and 
an electromagnetic calorimeter (ECL) based on CsI(Tl) crystals.
These detector elements are located inside a superconducting 
solenoid coil that provides a 1.5 T magnetic field. Muon identification
is provided by an array of resistive plate chambers (KLM) interspersed with
iron shielding that is used as the magnetic flux return.
For charged hadron identification, a likelihood ratio 
$\mathcal{L}^{}_{K} \equiv \mathcal{L}(K)/(\mathcal{L}(K) + \mathcal{L}(\pi))$
is formed based on $dE/dx$ measured in the CDC and the response of the 
ACC and TOF. Charged kaons are identified using a likelihood 
requirement that is about 86\% efficient for $K^\pm$ and has 
a $\pi^\pm$ misidentification rate of about~8\%.


We reconstruct the decay 
$D^{*\pm}\!\rightarrow\! D^{0}\pi^{\pm}_{s}$, 
$D^0\!\rightarrow\! K^\pm\pi^\mp\pi^+\pi^-$, in which the 
charge of the low-momentum (or ``slow'') pion $\pi^{\pm}_{s}$ 
identifies the flavor of the neutral $D$ candidate. For each event, 
the $D^0\rightarrow K^\pm\pi^\mp\pi^+\pi^-$ candidate is formed from 
combinations of four charged tracks.
We require that the likelihood ratio $\mathcal{L}^{}_{K}$ be greater than 0.7 for kaons and less than 0.4 for pions.
All track candidates are required to have a 
distance-of-closest-approach of less than 5.0~cm 
along the $z$ axis. In the transverse $r$-$\phi$ plane, 
we require a distance-of-closest-approach of less than 
2.0~cm for pion candidates and less than 1.0~cm for kaon 
candidates.
To suppress backgrounds from semileptonic decays, 
we reject tracks satisfying 
electron or muon identification criteria based on information from the ECL and KLM detectors.
This veto has an efficiency of 95\% for signal events and reduces the number of electron (muon) 
background events by 93\% (95\%).
We require that each track used to reconstruct 
the $D^{0}$ have at least two SVD hits in both the $r$-$\phi$ and 
$z$ coordinates. We retain events having a $K\pi\pi\pi$ invariant 
mass ($\mk3p$) satisfying $1.81$\gevm  $< \mk3p < 1.92$\gevm.

For $D^0\rightarrow K^+\pi^-\pi^+\pi^-$, when the momenta of a 
daughter kaon and pion are similar, their masses can be exchanged 
without a significant effect upon $\mk3p$. This misidentification 
leads to ``feed-through'' background from RS 
$D^0\rightarrow K^-\pi^+\pi^+\pi^-$ decays in the WS sample. 
To suppress this background, 
we recalculate $\mk3p$ of WS candidates after swapping the 
kaon and pion mass assignments and reject events in which 
$|\mk3p {\rm (swapped)} - m^{}_{D^0}| < 20$\gevm. 
From Monte Carlo (MC) simulation, we find that this veto has a 
signal efficiency of 92\% while rejecting 94\% of this background.

To suppress backgrounds from the singly Cabibbo-suppressed decay 
$D^{0}\rightarrow K^0_S\,K^{+}\pi^{-}$ followed by 
$K^0_S\rightarrow \pi^{+}\pi^{-}$, we veto events in which either
of the $\pi^{+}\pi^{-}$ daughter combinations has an invariant mass 
within 20\mevm\ ($3.3\sigma$ in resolution) of the $K^{0}_{S}$ mass.
This veto has an efficiency of 97\% for signal events and reduces 
the number of 
$K^{0}_{S}$ background events in Monte Carlo by 90\%.

To suppress background from random combinations of tracks, the daughter
tracks from the $D^{0}$ candidate are required to originate from a common 
vertex. To reconstruct the $D^*$ candidate, we perform a vertex fit that 
constrains the $D^0$ and the $\pi^{}_s$ candidate to the interaction point 
(IP) of the beams. 
The resolution on the mass difference
$Q \equiv M^{}_{\pi_{s}K3\pi} - \mk3p - m^{}_\pi$
is significantly improved by this requirement.
We require that the $\chi^{2}$ probability for each vertex 
fit be greater than 0.1\% and that $Q < 10$\mevm.
To eliminate $D$ mesons produced in $B\overline B$ events,
we require that the momentum of the $D^*$ candidate be 
greater than 2.5\gevp~in the center-of-mass (CM) frame.
After all selection requirements, the fraction of events 
containing multiple candidates is 8.6\%. For these events, 
we select the candidate 
that minimizes the sum of $\chi^{2}$ values divided by 
the sum of degrees of freedom (d.o.f.), where each sum
extends over both vertex fits.

We measure RS and WS signal yields by performing a two-dimensional 
binned maximum likelihood fit to the $\mk3p$ and $Q$ distributions. 
The signal and background probability density functions (PDFs) are 
determined from MC samples having sizes four times that of the 
data set.
Background PDF shapes are determined separately 
for RS and WS distributions and fixed in the fit.
The backgrounds are divided into four categories: 
(1) ``random~$\pi_{s}$,'' in which 
a CF $D^0\rightarrow K^-\pi^+\pi^+\pi^-$ decay is correctly 
reconstructed but is subsequently combined with
a random slow pion having the WS charge; 
(2) ``broken charm,'' in which a 
true $D^{*+}\rightarrow D^0\pi^+_s$ decay is combined with 
a misreconstructed $D^{0}$; 
(3) ``combinatoric,'' consisting of remaining backgrounds 
from $e^{+}e^{-}\rightarrow c\bar{c}$ production;
and 
(4) ``$uds$,'' consisting of combinatorial backgrounds from continuum 
$e^{+}e^{-}\rightarrow u\bar{u}, d\bar{d}, s\bar{s}$ production.
As no significant correlations are found between $\mk3p$ and $Q$
for the signal or backgrounds, we model each PDF as the product 
of one-dimensional functions.  
Background PDF shapes are parametrized in $Q$ using a threshold 
function of the form $Q^{1/2} + aQ^{3/2}$ for the random~$\pi_{s}$, 
combinatoric, and $uds$ components, and a broad Gaussian for the 
broken charm component. For $\mk3p$ a 
second-order Chebyshev polynomial  is used 
for the combinatoric and $uds$ components, and an ARGUS 
function~\cite{ARGUS} is used for the broken charm component.
The random~$\pi_{s}$ background is parametrized in $\mk3p$ using 
the same shape as used for the signal (see below).
We compare data and MC events in the sideband regions
$|Q-5.865$~\mevm$| > 2.0$
~\mevm\ for numerous kinematic distributions and find good agreement. 
These distributions include the $D^{*}$ momentum, 
the $\chi^2$ value of the vertex fits, particle identification 
likelihoods, the cosine of the angle between the $D^{0}$ and 
each of its daughter particles, and the momentum of each final 
state particle.
The background normalizations are floated in the fit.

\begin{figure}
\begin{center}
\vbox{
\hbox{\hspace{-0.1in}
\epsfig{file=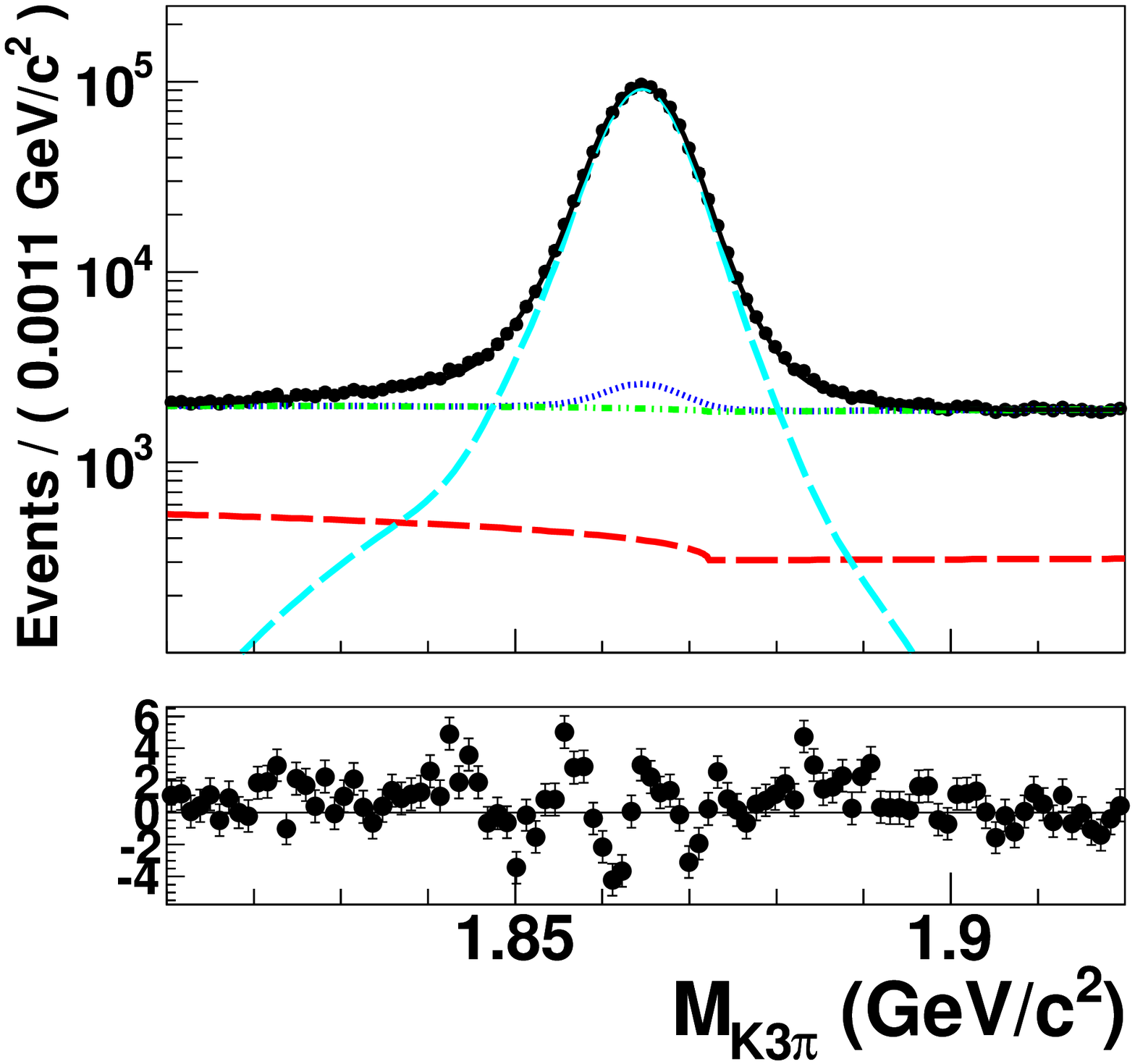,width=3.2in}
\epsfig{file=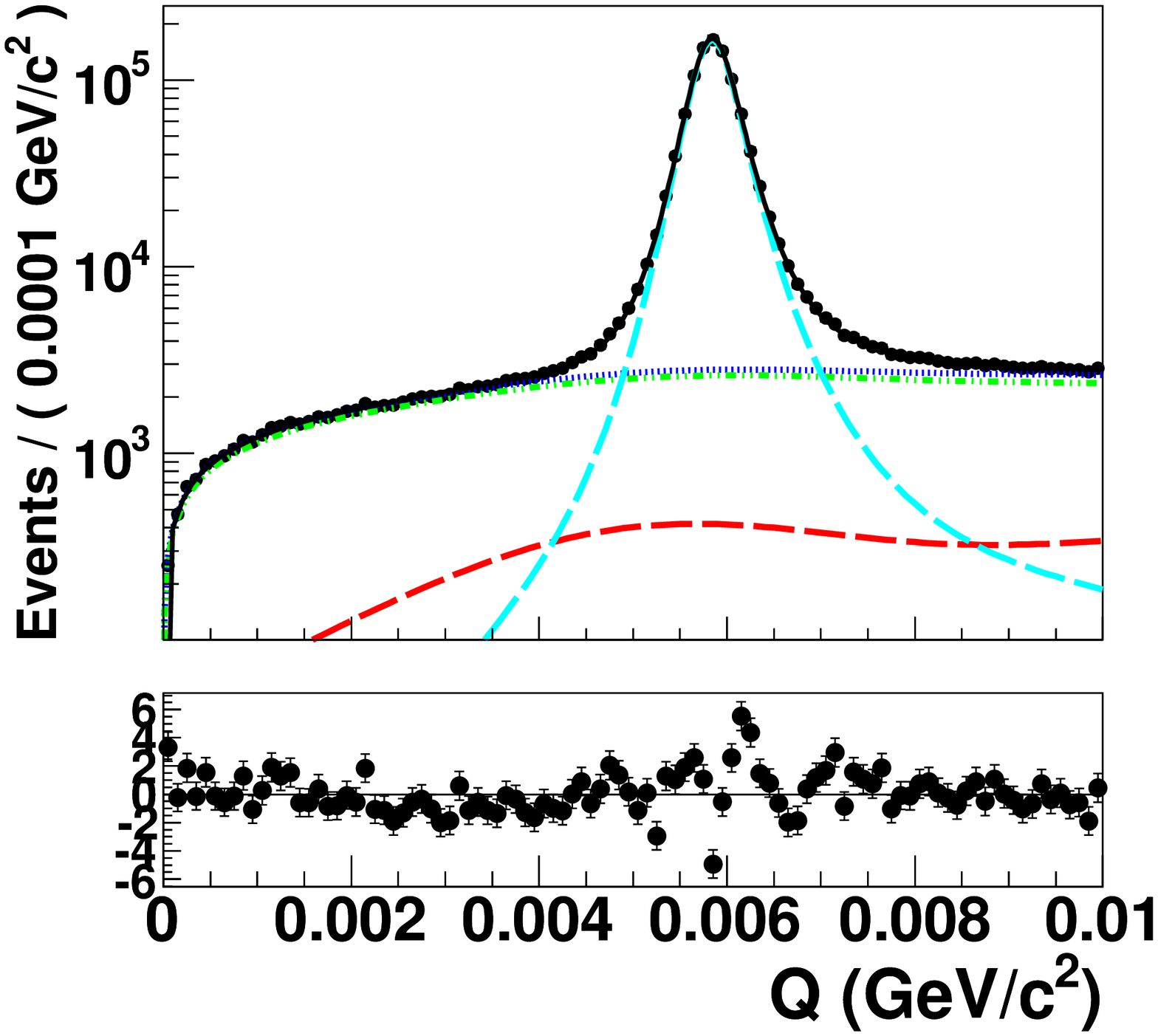,width=3.2in}
}
\hbox{\hspace{-0.1in}
\epsfig{file=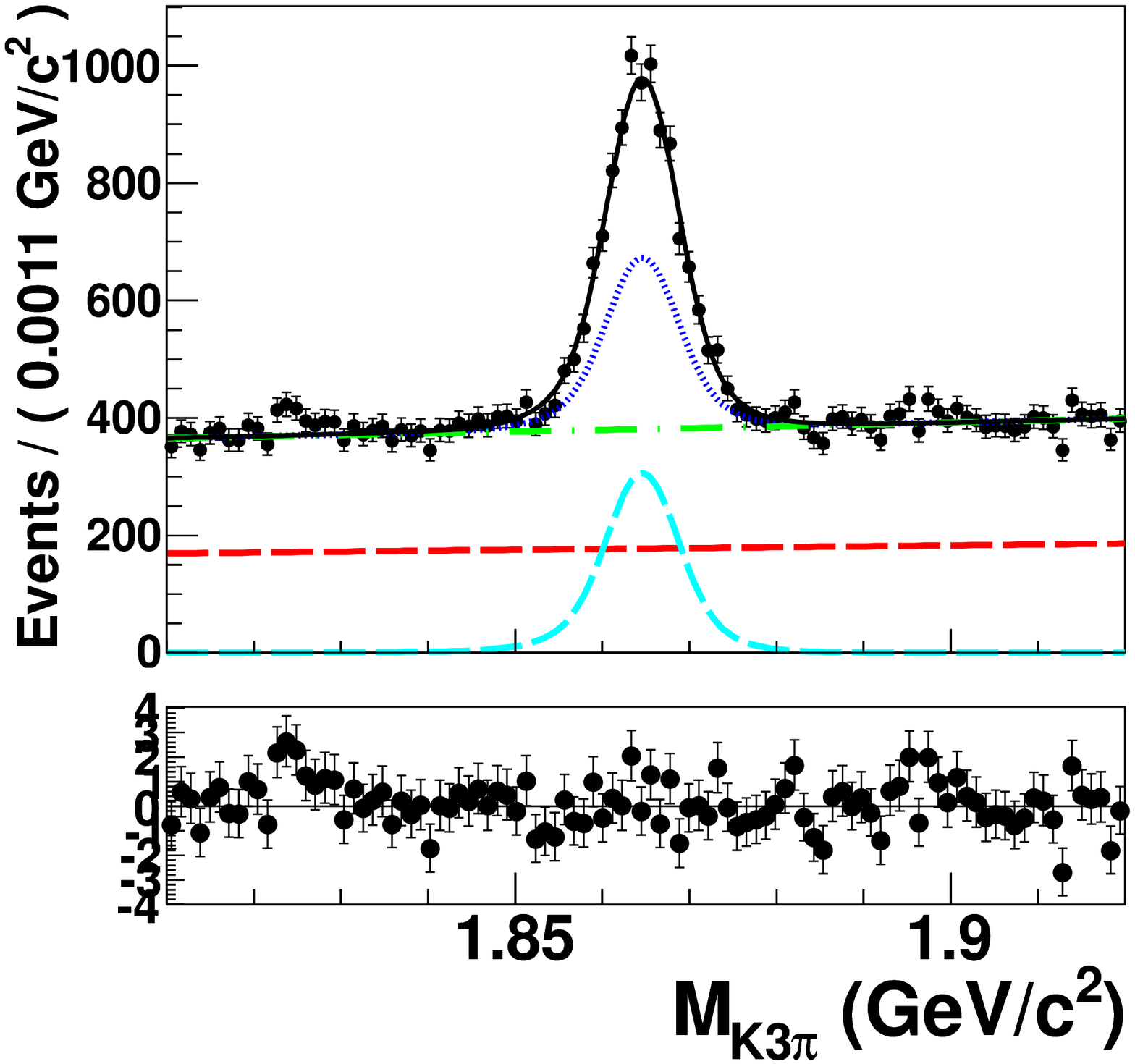,width=3.2in}
\epsfig{file=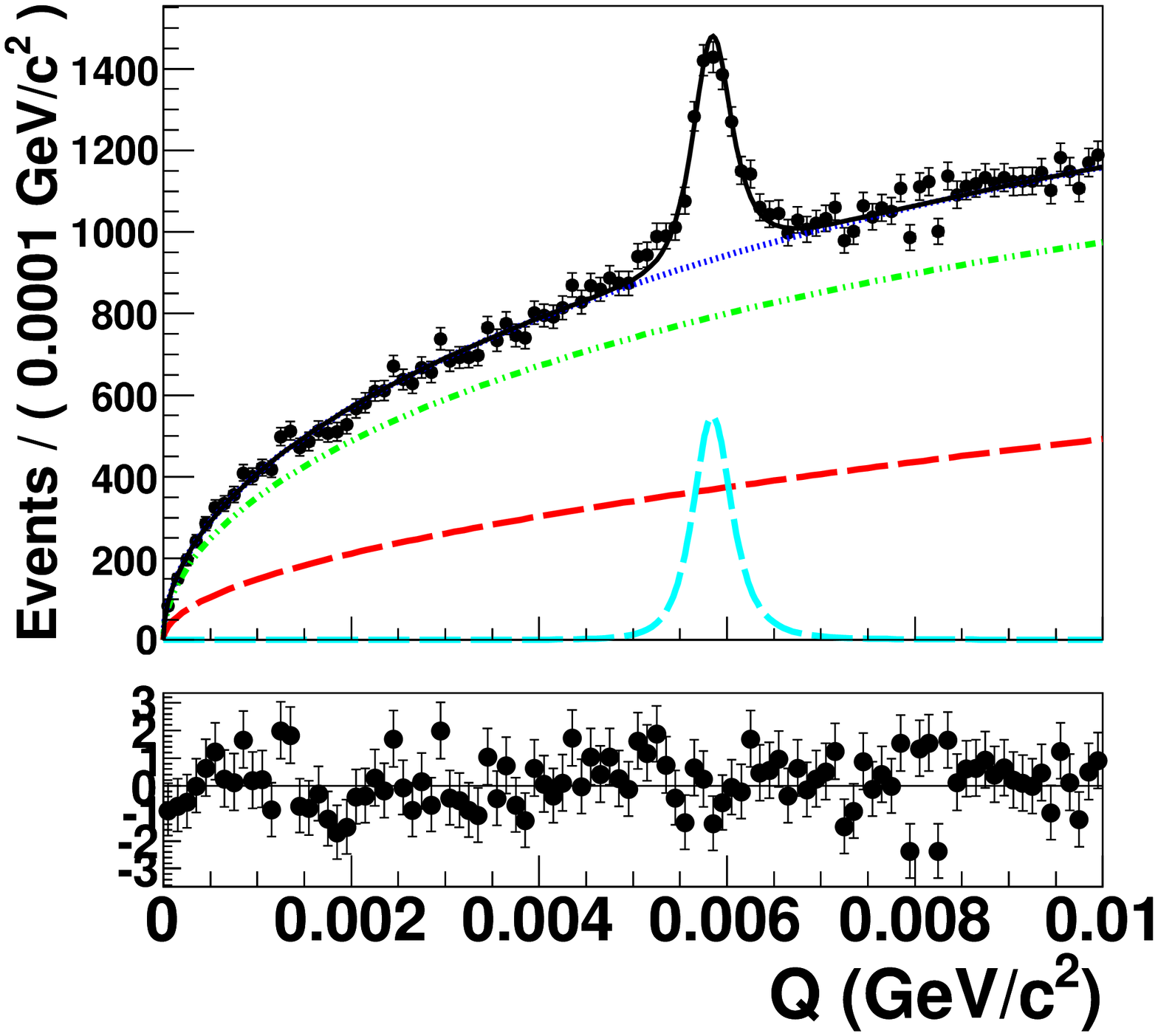,width=3.2in}
}
}
\end{center}
\vskip-0.30in
\caption{Fit projections for $\mk3p$ (left) and $Q$ (right). 
The top (bottom) row shows $\cf$ ($\dcs$) candidates.
Events plotted for $\mk3p$ are required to lie in 
the signal region for $Q$, and vice versa.
The peaking dashed curves show the signal PDF (cyan);
the non-peaking dashed curves show broken charm and $uds$ backgrounds (red); 
the dash-dotted curves include combinatoric backgrounds (green);
and the dotted curves include random $\pi^{}_{s}$ backgrounds (blue).
The fit residuals $(N_{\rm o}-N_{\rm p})/\sqrt{N_{\rm o}}$ are plotted below each fit projection, where $N_{\rm o}$ ($N_{\rm p}$) is the observed (predicted) event yield.}
\label{fig:fit_results}
\end{figure}

The RS signal PDF is parametrized in $\mk3p$ as the sum of one Gaussian and two 
bifurcated Gaussians with a common mean, and in $Q$ as the sum of a bifurcated 
Student's $t$-distribution and a bifurcated Gaussian with common mean.
For both distributions, the relative fraction between the single Gaussian 
and the remaining function(s) is fixed to the value obtained from the MC 
while all other parameters are floated in the fit. The RS signal PDF is 
used also for the WS signal PDF.
Since the WS and RS samples are fitted simultaneously, the ratio 
of WS to RS signal yields is extracted directly from the fit.
We obtain a RS yield of $990594\pm 1901$ events and a ``raw'' ratio 
of WS to RS yields of $\rawrws = (0.339 \pm 0.008)\%$.
This value must be corrected for the ratio of 
overall efficiencies of RS and WS decays.
Projections of the fit are shown in Fig.~\ref{fig:fit_results}. 
The fitted RS yield and $\rawrws$ value correspond to a 
WS yield of $3358\pm 79$ events. The goodness of fit
is satisfactory: for WS (RS) decays,
$\chi^2/{\rm d.o.f.}= 1.17\,(1.89)$ for $\mk3p$ and 0.90\,(1.43) for $Q$.


As $\dcs$ and $\cf$ decays proceed largely through intermediate 
resonances, RS and WS events are expected to 
have different distributions across the phase space.
If the detector acceptance and reconstruction efficiencies 
vary over phase space,
the overall efficiencies for RS and WS decays will differ from each other.
The ratio of these efficiencies is needed to correct~$\rawrws$.
 
To obtain the ratio of efficiencies, we divide RS and WS events 
into 576 bins in a five-dimensional phase space.
These dimensions consist of the invariant mass combinations for 
$K^{\pm}\pi^{\pm}$, $K^{\pm}\pi^{\mp}_{1}$,  $K^{\pm}\pi^{\mp}_{2}$, 
$\pi^{\pm}\pi^{\mp}_{1}$, and $\pi^{\pm}\pi^{\mp}_{2}$, where
$\pi^{}_{1}$ and $\pi^{}_{2}$ label the pions with same sign charge,
and $|p_{\pi_{1}}| > |p_{\pi_{2}}|$.
 The binning is chosen to correspond to the 
structure present in these variables. 
The efficiency for each bin ($\epsilon^{}_{i}$) is obtained 
using MC. We estimate background in the data for bin $i$ by 
multiplying the total background yield ($N_{\rm bkg}$) by the 
fraction of background events in that bin ($f^{}_i$) as obtained 
from MC simulation. The total background yield is determined 
from a two-dimensional fit to the $\mk3p$-$Q$ distribution
of data. The total signal yield is calculated as
\begin{eqnarray}
N'(K\pi\pi\pi) & = & \sum_{i=1}^{576} \frac{N^{}_{i} - 
N_{\rm bkg} \cdot f_{i}}{\epsilon_{i}}\,,
\label{eq:eff_correct_yield}
\end{eqnarray}
where $N^{}_{i}$ is the number of candidate events in bin $i$. 
The reconstruction efficiency for either $\dcs$ or $\cf$ decays 
is calculated as
\begin{eqnarray}
\epsilon (K\pi\pi\pi) & = & \frac{1}{N'}\sum_{i=1}^{576} 
\left(N^{}_{i} - N_{\rm bkg} \cdot f_{i}\right)\,,
\label{eq:efficiency}
\end{eqnarray}
and thus
\begin{equation}
\rws = \rawrws\cdot\frac{\epsilon(K^-\pi^+\pi^+\pi^-)}{\epsilon(K^+\pi^-\pi^+\pi^-)}
 = \frac{N'(K^+\pi^-\pi^+\pi^-)}{N'(K^-\pi^+\pi^+\pi^-)}\,.
\label{eq:effic_correct_rws}
\end{equation}
Only events located within a signal region 
$|m_{K3\pi}-m_{D^{0}}| < 0.01$~GeV and $|Q - Q_{0}| < 0.002$~GeV/$c^2$
are used to detemine the efficiency correction. 
The efficiency-corrected yields are 
$N'(K^+\pi^-\pi^+\pi^-) = 37297 \pm 881$ and 
$N'(K^-\pi^+\pi^+\pi^-) = (1.151 \pm 0.002) \times10^{7} $; 
thus $\rws = (0.324 \pm 0.008)$\%.


We consider various sources of systematic uncertainty as
listed in Table~\ref{tab:systematics}. Since we measure the
ratio of topologically similar RS and WS decays, many
systematic uncertainties cancel.

To determine the systematic uncertainty associated with the
ratio of efficiencies, we propagate the statistical errors for 
$\epsilon_{i}$ and $f_{i}$ via a Monte Carlo method
as follows. We generate values for $\epsilon_{i}$
and $f_{i}$ in all 576 bins. These values are sampled from
Gaussian distributions having mean values equal to the nominal parameter
values and standard deviations equal to their uncertainties. 
We then recalculate $\rws$ using these sampled values. 
We repeat this procedure $10^5$ times and plot the resulting 
distribution of $\rws$. 
The RMS of this distribution ($\pm 0.0041$) is taken as the 
systematic error associated with the efficiency correction.

To estimate the contribution associated with event selection
criteria, we vary each selection criterion over a suitable 
range and remeasure $\rws$ for each variation. The identification 
likelihood ratio $\mathcal{L}^{}_{K}$ is varied over the range 
0.5--0.9 for kaon candidates and 0.1--0.5 for pion candidates.
The momentum requirement for $D^{*}$ candidates is varied 
over the range 2.3--2.7~\gevp. For each selection criterion, 
the largest positive and negative deviation of $\rws$ from 
the nominal value is taken as the systematic error.
The error due to multiple candidates is obtained by removing 
all events containing multiple candidates (8.6\% of events) 
and refitting for $\rws$; the deviation observed is taken 
as the error. To determine the uncertainty associated with background 
PDF shapes (which are taken from MC and differ for RS and WS 
events), we vary the parameters of each background PDF by 
$\pm 1\sigma$, 
where $\sigma$ corresponds to the statistical error
from the fit to MC. 
For each variation, the data is refit and the deviation of 
$\rws$ from the nominal value  is recorded.
The uncertainty due to a given background 
PDF is taken as the sum in quadrature of all deviations
observed when varying the individual parameters.
The systematic error due to uncertainty in the
signal PDF is negligibly small.
To check for possible bias in our fit results, we repeat 
the fit for Monte Carlo samples
(each corresponding to the size of the data set) having
different values of $\rws$. Comparing the fit results for
$\rws$ with the true values shows no visible fit bias.
The total systematic error is taken to be the sum in
quadrature of all individual contributions. 
Our final result is
\begin{eqnarray} 
\rws & = & (0.324 \pm 0.008 \pm 0.007)\%.
\end{eqnarray}
Multiplying this value by the well-measured RS branching fraction 
${\cal B}(D^0\rightarrow K^-\pi^+\pi^+\pi^-)= (8.07\,^{+0.21}_{-0.19})$\%~\cite{pdg_2012}
gives a WS branching fraction
\begin{eqnarray}
{\cal B}(D^0\!\rightarrow\! K^+\pi^-\pi^+\pi^-) & = & 
(2.61 \pm 0.06\,^{+0.09}_{-0.08}) \!\times \! 10^{-4}.~~~~~
\end{eqnarray} 
By combining our measurement of $\rws$ with world average 
values~\cite{HFAG_avg} for $x$ and $y$, and recent 
measurements~\cite{Lowrey} of $\alpha$ and $\delta$,
we extract $R_{D}$ from  Eq.~\ref{eq:rws}. We use a MC 
method to propagate the errors for the parameters and
obtain $R_{D} = (0.327^{+0.019}_{-0.016})\%$.

\begin{table}[ptbh]
\caption{Summary of systematic errors for $\rws$.
The total systematic error is obtained by summing 
all contributions in quadrature.}
\begin{center}
\renewcommand{\arraystretch}{0.9}
\begin{tabular*}{0.47\textwidth}
{@{\extracolsep{\fill}} l | c c }
\hline \hline
Source & $+\Delta R$ (\%) & $-\Delta R$ (\%)  \\
\hline 
  Kaon ID  & 0.0008  &  0.0006   \\
  Pion ID  &  0.0003  &   0.0024  \\
  $D^*$ Momentum  &  0.0029    &  0.0037 \\
\hline
  Multiple Candidates & 0.0024   &  0.0024  \\
\hline
  $uds$  &  0.0012  &  0.0002   \\
  Combinatoric  &   0.0034   &  0.0025  \\
  Slow $\pi$  &  0.0009   &  0.0003 \\
  Broken  &   0.0010  &  0.0008  \\
\hline
  Efficiency Correction  &  0.0041   &  0.0041  \\
\hline
  Sum  &  0.0069   &   0.0070   \\
\hline\hline
\end{tabular*} 
\end{center}
\label{tab:systematics}
\end{table}


In summary, we have measured the wrong-sign ratio 
$\rws=\Gamma(\dcs)/\Gamma(\cf)$ using $e^{+}e^{-}$ data 
collected at or near the $\Upsilon(4S)$ resonance.
After correcting for differences in reconstruction efficiencies 
between RS and WS events, we obtain 
$\rws = (0.324 \pm 0.008 \pm 0.007)$\%,
where the first uncertainty is statistical and the second is systematic.
This is the most precise measurement of $\rws$ to date. 
Using a MC method to extract $R_{D}$ from  Eq.~\ref{eq:rws},
we obtain $R_{D} = (0.327^{+0.019}_{-0.016})\%$.
Multiplying $\rws$ by the branching fraction for 
$D^0\rightarrow K^-\pi^+\pi^+\pi^-$ gives
${\cal B}(D^0\!\rightarrow\! K^+\pi^-\pi^+\pi^-) = 
(2.61 \pm 0.06\,^{+0.09}_{-0.08}) \!\times \! 10^{-4}$.
This result is substantially more precise than the current PDG value 
of $(2.61\,^{+0.21}_{-0.19}) \! \times \! 10^{-4}$~\cite{pdg_2012}.


We thank the KEKB group for the excellent operation of the
accelerator; the KEK cryogenics group for the efficient
operation of the solenoid; and the KEK computer group,
the National Institute of Informatics, and the 
PNNL/EMSL computing group for valuable computing
and SINET4 network support.  We acknowledge support from
the Ministry of Education, Culture, Sports, Science, and
Technology (MEXT) of Japan, the Japan Society for the 
Promotion of Science (JSPS), and the Tau-Lepton Physics 
Research Center of Nagoya University; 
the Australian Research Council and the Australian 
Department of Industry, Innovation, Science and Research;
Austrian Science Fund under Grant No. P 22742-N16;
the National Natural Science Foundation of China under
contract No.~10575109, 10775142, 10875115 and 10825524; 
the Ministry of Education, Youth and Sports of the Czech 
Republic under contract No.~MSM0021620859;
the Carl Zeiss Foundation, the Deutsche Forschungsgemeinschaft
and the VolkswagenStiftung;
the Department of Science and Technology of India; 
the Istituto Nazionale di Fisica Nucleare of Italy; 
The BK21 and WCU program of the Ministry Education Science and
Technology, National Research Foundation of Korea Grant No.\ 
2010-0021174, 2011-0029457, 2012-0008143, 2012R1A1A2008330,
BRL program under NRF Grant No. KRF-2011-0020333,
and GSDC of the Korea Institute of Science and Technology Information;
the Polish Ministry of Science and Higher Education and 
the National Science Center;
the Ministry of Education and Science of the Russian
Federation and the Russian Federal Agency for Atomic Energy;
the Slovenian Research Agency;
the Basque Foundation for Science (IKERBASQUE) and the UPV/EHU under 
program UFI 11/55;
the Swiss National Science Foundation; the National Science Council
and the Ministry of Education of Taiwan; and the U.S.\
Department of Energy and the National Science Foundation.
This work is supported by a Grant-in-Aid from MEXT for 
Science Research in a Priority Area (``New Development of 
Flavor Physics''), and from JSPS for Creative Scientific 
Research (``Evolution of Tau-lepton Physics'').

\end{document}

%% file: authors_pub398.tex
\noaffiliation
\affiliation{University of the Basque Country UPV/EHU, 48080 Bilbao}
\affiliation{University of Bonn, 53115 Bonn}
\affiliation{Budker Institute of Nuclear Physics SB RAS and Novosibirsk State University, Novosibirsk 630090}
\affiliation{Faculty of Mathematics and Physics, Charles University, 121 16 Prague}
\affiliation{University of Cincinnati, Cincinnati, Ohio 45221}
\affiliation{Deutsches Elektronen--Synchrotron, 22607 Hamburg}
\affiliation{Justus-Liebig-Universit\"at Gie\ss{}en, 35392 Gie\ss{}en}
\affiliation{II. Physikalisches Institut, Georg-August-Universit\"at G\"ottingen, 37073 G\"ottingen}
\affiliation{Hanyang University, Seoul 133-791}
\affiliation{University of Hawaii, Honolulu, Hawaii 96822}
\affiliation{High Energy Accelerator Research Organization (KEK), Tsukuba 305-0801}
\affiliation{Ikerbasque, 48011 Bilbao}
\affiliation{Indian Institute of Technology Guwahati, Assam 781039}
\affiliation{Indian Institute of Technology Madras, Chennai 600036}
\affiliation{Institute of High Energy Physics, Chinese Academy of Sciences, Beijing 100049}
\affiliation{Institute of High Energy Physics, Vienna 1050}
\affiliation{Institute for High Energy Physics, Protvino 142281}
\affiliation{INFN - Sezione di Torino, 10125 Torino}
\affiliation{Institute for Theoretical and Experimental Physics, Moscow 117218}
\affiliation{J. Stefan Institute, 1000 Ljubljana}
\affiliation{Kanagawa University, Yokohama 221-8686}
\affiliation{Institut f\"ur Experimentelle Kernphysik, Karlsruher Institut f\"ur Technologie, 76131 Karlsruhe}
\affiliation{Korea Institute of Science and Technology Information, Daejeon 305-806}
\affiliation{Korea University, Seoul 136-713}
\affiliation{Kyungpook National University, Daegu 702-701}
\affiliation{\'Ecole Polytechnique F\'ed\'erale de Lausanne (EPFL), Lausanne 1015}
\affiliation{Faculty of Mathematics and Physics, University of Ljubljana, 1000 Ljubljana}
\affiliation{University of Maribor, 2000 Maribor}
\affiliation{Max-Planck-Institut f\"ur Physik, 80805 M\"unchen}
\affiliation{School of Physics, University of Melbourne, Victoria 3010}
\affiliation{Moscow Physical Engineering Institute, Moscow 115409}
\affiliation{Graduate School of Science, Nagoya University, Nagoya 464-8602}
\affiliation{Kobayashi-Maskawa Institute, Nagoya University, Nagoya 464-8602}
\affiliation{Nara Women's University, Nara 630-8506}
\affiliation{National Central University, Chung-li 32054}
\affiliation{National United University, Miao Li 36003}
\affiliation{Department of Physics, National Taiwan University, Taipei 10617}
\affiliation{H. Niewodniczanski Institute of Nuclear Physics, Krakow 31-342}
\affiliation{Nippon Dental University, Niigata 951-8580}
\affiliation{Niigata University, Niigata 950-2181}
\affiliation{University of Nova Gorica, 5000 Nova Gorica}
\affiliation{Osaka City University, Osaka 558-8585}
\affiliation{Pacific Northwest National Laboratory, Richland, Washington 99352}
\affiliation{Panjab University, Chandigarh 160014}
\affiliation{University of Pittsburgh, Pittsburgh, Pennsylvania 15260}
\affiliation{University of Science and Technology of China, Hefei 230026}
\affiliation{Seoul National University, Seoul 151-742}
\affiliation{Soongsil University, Seoul 156-743}
\affiliation{Sungkyunkwan University, Suwon 440-746}
\affiliation{School of Physics, University of Sydney, NSW 2006}
\affiliation{Tata Institute of Fundamental Research, Mumbai 400005}
\affiliation{Excellence Cluster Universe, Technische Universit\"at M\"unchen, 85748 Garching}
\affiliation{Toho University, Funabashi 274-8510}
\affiliation{Tohoku Gakuin University, Tagajo 985-8537}
\affiliation{Tohoku University, Sendai 980-8578}
\affiliation{Department of Physics, University of Tokyo, Tokyo 113-0033}
\affiliation{Tokyo Institute of Technology, Tokyo 152-8550}
\affiliation{Tokyo Metropolitan University, Tokyo 192-0397}
\affiliation{Tokyo University of Agriculture and Technology, Tokyo 184-8588}
\affiliation{University of Torino, 10124 Torino}
\affiliation{CNP, Virginia Polytechnic Institute and State University, Blacksburg, Virginia 24061}
\affiliation{Wayne State University, Detroit, Michigan 48202}
\affiliation{Yamagata University, Yamagata 990-8560}
\affiliation{Yonsei University, Seoul 120-749}
  \author{E.~White}\affiliation{University of Cincinnati, Cincinnati, Ohio 45221} 
  \author{A.~J.~Schwartz}\affiliation{University of Cincinnati, Cincinnati, Ohio 45221} 
  \author{I.~Adachi}\affiliation{High Energy Accelerator Research Organization (KEK), Tsukuba 305-0801} 
  \author{H.~Aihara}\affiliation{Department of Physics, University of Tokyo, Tokyo 113-0033} 
  \author{D.~M.~Asner}\affiliation{Pacific Northwest National Laboratory, Richland, Washington 99352} 
  \author{V.~Aulchenko}\affiliation{Budker Institute of Nuclear Physics SB RAS and Novosibirsk State University, Novosibirsk 630090} 
  \author{T.~Aushev}\affiliation{Institute for Theoretical and Experimental Physics, Moscow 117218} 
  \author{A.~M.~Bakich}\affiliation{School of Physics, University of Sydney, NSW 2006} 
  \author{A.~Bala}\affiliation{Panjab University, Chandigarh 160014} 
  \author{V.~Bhardwaj}\affiliation{Nara Women's University, Nara 630-8506} 
% ...
\collaboration{The Belle Collaboration}

%% file: K3Pi.bbl
\begin{thebibliography}{99}
\bibitem{cc}
The inclusion of charge-conjugated modes is implied throughout this paper.
\end{thebibliography}
